\documentclass[epj,twocolumns]{svjour_mod}
\usepackage{amsmath,graphicx,slashed,color}
\usepackage{graphicx,epsfig,amssymb,bm}
\usepackage[utf8]{inputenc}
\def \beq{\begin{equation}}
\def \eeq{\end{equation}}
\def \beqa{\begin{eqnarray}}
\def \eeqa{\end{eqnarray}}

\usepackage{ragged2e}

\usepackage{booktabs}

\usepackage{tabularx}
\newcolumntype{C}[1]{>{\centering\arraybackslash}p{#1}}

\usepackage{graphicx}
\usepackage{url}
\usepackage{bm}
\usepackage{hyperref}
\usepackage{amsmath}
\usepackage{amssymb}
\usepackage{listings}
\usepackage{color}

\usepackage{tabularx}
\newcolumntype{C}[1]{>{\centering\arraybackslash}p{#1}}

\def\bea{\begin{eqnarray}}
\def\eea{\end{eqnarray}}

\def\gsim{\mathrel{\rlap{\lower4pt\hbox{\hskip1pt$\sim$}}
    \raise1pt\hbox{$>$}}}         
\def\lsim{\mathrel{\rlap{\lower4pt\hbox{\hskip1pt$\sim$}}
    \raise1pt\hbox{$<$}}}         
\def\smallfrac#1#2{\hbox{${{#1}\over{#2}}$}}
\def\half{\smallfrac{1}{2}}

\def\smallfrac#1#2{\hbox{$\frac{#1}{#2}$}}
\newcommand{\be}{\begin{equation}}
\newcommand{\ee}{\end{equation}}
\newcommand{\bi}{\begin{itemize}}
\newcommand{\ei}{\end{itemize}}
\newcommand{\ben}{\begin{enumerate}}
\newcommand{\een}{\end{enumerate}}

\newcommand{\lp}{\left(}
\newcommand{\rp}{\right)}

\begin{document}
\title{A First Determination of Parton Distributions with Theoretical Uncertainties}

\PACS{{}{}}

\author{
Rabah Abdul Khalek\inst{1,2}
\and Richard D. Ball\inst{3}
\and Stefano Carrazza\inst{4}
\and Stefano Forte\inst{4}
\and Tommaso Giani\inst{3}
\and Zahari Kassabov\inst{5}
\and Emanuele R. Nocera\inst{6}
\and Rosalyn L. Pearson\inst{3}
\and Juan Rojo\inst{6,7}
\and Luca Rottoli\inst{8,9}
\and Maria Ubiali\inst{10}
\and Cameron Voisey\inst{5}
\and Michael Wilson\inst{3}
}

\institute{Department of Physics and Astronomy, VU Amsterdam, De Boelelaan 1081, NL-1081, HV Amsterdam, The Netherlands
\and
Nikhef, Science Park 105, NL-1098 XG Amsterdam, The Netherlands
\and
The Higgs Centre for Theoretical Physics, University of Edinburgh,
  JCMB, KB, Mayfield Rd, Edinburgh EH9 3JZ, Scotland
\and
Tif Lab, Dipartimento di Fisica, Universit\`a di Milano and
  INFN, Sezione di Milano, Via Celoria 16, I-20133 Milano, Italy
\and
Cavendish Laboratory, University of Cambridge, Cambridge, CB3
0HE, United Kingdom
\and
Nikhef, Science Park 105, NL-1098 XG Amsterdam, The Netherlands
\and
Department of Physics and Astronomy, VU Amsterdam, De Boelelaan 1081, NL-1081, HV Amsterdam, The Netherlands
\and
Dipartimento di Fisica G. Occhialini, U2, Università degli Studi di Milano-Bicocca,
  Piazza della Scienza, 3, 20126 Milano, Italy
\and
INFN, Sezione di Milano-Bicocca, 20126, Milano, Italy
\and
DAMTP, University of Cambridge, Wilberforce Road, Cambridge, CB3 0WA, United Kingdom
}


\date{\today}

\abstract{
  The parton distribution functions (PDFs) which characterize the
  structure of the proton
  are currently one of the dominant sources of uncertainty in the predictions for
  most  processes measured at the Large Hadron Collider (LHC).
  Here we present the first extraction of the proton PDFs that accounts for
  the missing higher order uncertainty
  (MHOU) in the fixed-order QCD calculations used in PDF determinations.
  We demonstrate that the MHOU can be included as a contribution to
  the covariance  matrix used for the PDF fit, and then introduce
  prescriptions for the computation of this covariance matrix using
  scale variations.
  We validate our results at next-to-leading order (NLO) by comparison to
  the known next order (NNLO) corrections.
  We then construct variants of the NNPDF3.1 NLO PDF set
  that include the effect of the MHOU, and assess their impact
  on the central values and uncertainties of the resulting PDFs.
}

\maketitle

The search for new physics at present~\cite{deFlorian:2016spz}
and future~\cite{Cepeda:2019klc} high-energy colliders,
and specifically at the LHC, has turned from the mapping of the
energy frontier to the exploration of the precision frontier: looking for subtle deviations from Standard Model predictions.
In this endeavor, an accurate estimate of uncertainties
associated with these predictions is crucial.
At present, these uncertainties have two main origins.
The first is the
missing higher order uncertainty (MHOU)
from the truncation of the QCD perturbative expansion.
The second  is related to knowledge of the structure
of the colliding protons, as encoded in the parton distributions
(PDFs)~\cite{Gao:2017yyd}.

PDFs are extracted by comparing theoretical predictions to experimental data.
Currently, PDF uncertainties only
account for the propagated statistical and
systematic errors on the measurements used in their
determination.
However, the same MHOU
which affects predictions at the LHC also affect predictions for the
various processes that enter the PDF determination.
These are
currently neglected, perhaps because they are believed to be
generally less important than experimental uncertainties.
However, as PDFs become more
precise, in particular thanks to ever tighter constraints
from LHC data~\cite{Rojo:2015acz}, MHOUs in
PDF determinations will eventually become significant.
Already in recent PDF sets making extensive
use of LHC data, such as NNPDF3.1~\cite{Ball:2017nwa},
the shift between PDFs at next-to-leading order (NLO) and the next
order (NNLO) is sometimes
larger than the PDF uncertainties from the experimental data.

Here we present the first PDF extraction that systematically accounts
for the MHOU in the QCD calculations used to extract them.
MHOUs are routinely estimated by varying the arbitrary renormalization
$\mu_r$ and factorization $\mu_f$ scales of perturbative
computations~\cite{deFlorian:2016spz}, though alternative methods
have also been proposed~\cite{Cacciari:2011ze,David:2013gaa,Bagnaschi:2014wea}.
Our inclusion of the MHOU in a PDF fit involves two steps:
first we establish how theoretical uncertainties can be included
in such a fit through a
covariance matrix~\cite{Ball:2018odr,Ball:2018twp}, and then we  find
a way of computing and validating the covariance matrix associated with the
MHOU using scale variations~\cite{Pearson:2018tim}.
By producing variants of NNPDF3.1 which include the MHOU, we are
then able to finally address the long-standing question of their
impact on state-of-the-art PDF sets.
A detailed discussion of our results is presented in a companion
paper~\cite{AbdulKhalek:2019ihb}, to which we refer for full
computational details, definitions, proofs and results.

Assuming that theory uncertainties can be modeled as Gaussian distributions,
in the same way as experimental systematics, then the associated
theory covariance matrix $S_{ij}$ can be expressed in terms of nuisance parameters
\be
\label{eq:covth}
S_{ij} = \smallfrac{1}{N}\sum_k \Delta_i^{(k)}\Delta_j^{(k)} \, ,
\ee
where $\Delta_i^{(k)}=T^{(k)}_i-T_i^{(0)}$ is the expected shift with respect
to the central theory prediction for the $i$-th cross-section,
$T_i^{(0)}$, due to the theory uncertainty, and $N$ is a normalization
factor determined by the number of independent nuisance parameters.
Since theory uncertainties are independent of the
experimental ones, the two can be
combined in quadrature:
the $\chi^2$ used to assess the agreement of theory and data is
given by
\be
\label{eq:chi2}
\chi^2=\sum_{i,j=1}^{N_{\rm dat}}\lp D_i-T_i^{(0)}\rp
\lp S+C\rp^{-1}_{ij} \lp  D_j-T_j^{(0)}\rp \, ,
\ee
with $D_i$ the central experimental value of the $i$-th datapoint,
and $C_{ij}$ the experimental covariance matrix.
More details of the implementation of the theory covariance matrix in PDF
fits may be found in Refs.~\cite{Ball:2018odr,Ball:2018twp}.

The choice of nuisance parameters $\Delta_i^{(k)}$ used in
Eq.~(\ref{eq:covth}) to estimate a particular theoretical uncertainty is
not unique, reflecting the fact that such estimates
always have some degree of arbitrariness.
Here we focus on the MHOU, and choose to use
scale variations to estimate $\Delta_i^{(k)}$.
A standard
procedure~\cite{deFlorian:2016spz}
is the so-called 7-point prescription, in which
the MHOU is estimated from the envelope of results obtained with
the following scales
\be
\nonumber
\lp k_f, k_r\rp \in \{ (1,1),(2,2),(\half,\half),(2,1),(1,2),(\half,1),(1,\half) \}
\ee
where $k_{r}=\mu_{r}/\mu_{r}^{(0)}$ and $k_{f}=\mu_{f}/\mu_{f}^{(0)}$  are the ratios
of the renormalization and factorization scales to their central values.
Varying $\mu_r$ estimates the MHOU in the hard coefficient function of
the specific process, while the $\mu_f$ variation estimates the MHOU
in PDF evolution.

In order to compute a covariance matrix, we must not only choose a set of
scale variations, but also make some assumptions about the way
they are correlated.
We do this by, first of all, classifying
the input datasets used in PDF fits
into processes as indicated in
Table~\ref{eq:expclassification}: charged-current (CC) and neutral-current (NC) deep-inelastic scattering (DIS), Drell-Yan (DY) production of gauge bosons (invariant mass, transverse momentum, and rapidity distributions), single-jet inclusive and top pair production cross-sections.
Note that this step
requires making an educated guess as to which cross-sections are likely to have
a similar structure of higher-order corrections.

\begin{table}[t]
  \centering
  \renewcommand*{\arraystretch}{1.3}
  \begin{tabular}{|c|c|}
    \hline
    Process Type  & Datasets \\
    \hline
    DIS NC  &   NMC, SLAC, BCDMS, HERA NC \\
    DIS CC  &   NuTeV, CHORUS, HERA CC \\
    DY  & CDF, D0, ATLAS, CMS, LHCb ($y$, $p_T$, $M_{ll}$) \\
    JET  & ATLAS, CMS inclusive jets \\
    TOP  & ATLAS, CMS total+differential cross-sections \\
    \hline
  \end{tabular}
  \caption{\label{eq:expclassification}
   Classification of  datasets into  process types.
  }
\end{table}

Next, we formulate a variety of prescriptions for how to construct
Eq.~(\ref{eq:covth}) by picking a set of scale variations and
correlation patterns.
A simple possibility is the 3-point prescription, in which we vary 
both scales coherently (thus setting $k_f=k_r$) by a fixed amount about the central
value, independently for each process.
More sophisticated  prescriptions vary
the two scales independently, but by the same amount, and assume that while $\mu_r$ is
only correlated within a given process, $\mu_f$ is fully
correlated among processes.
This assumption is based on the observation that
$\mu_f$ variations estimate the MHOU in the evolution equations,
which are universal (process-independent), though it is an
approximation given that the evolution of different PDFs is governed by
different anomalous dimensions, which do not necessarily share the
same MHO corrections.

We then proceed to the validation of the resulting covariance matrices at
NLO.
We use the same experimental data and theory calculations
as in the NNPDF3.1 $\alpha_s$
study~\cite{Ball:2018iqk} with two minor differences:
the value of the lower kinematic
cut has been increased from
$Q_{\rm min}^2=2.69$~GeV$^2$ to
$13.96$~GeV$^2$ in order to ensure the validity of the
perturbative QCD expansion when scales
are varied downwards, and the HERA $F_2^b$ and
fixed-target Drell-Yan cross-sections have been removed, for technical
reasons related to difficulties in implementing scale variation.
In total we then have $N_{\rm dat}=2819$ data points.
The theory covariance matrix $S_{ij}$ has been constructed
by means of the {\tt ReportEngine} software~\cite{zahari_kassabov_2019_2571601}
taking as input the
scale-varied NLO theory cross-sections $T_i(k_f,k_r)$, provided
by {\tt APFEL}~\cite{Bertone:2013vaa}
for the DIS structure functions
and by {\tt APFELgrid}~\cite{Bertone:2016lga} combined with
{\tt APPLgrid}~\cite{Carli:2010rw} for the hadronic
cross-sections.

\begin{figure}[t]
    \begin{center}
    \makebox{\includegraphics[width=0.99\columnwidth]{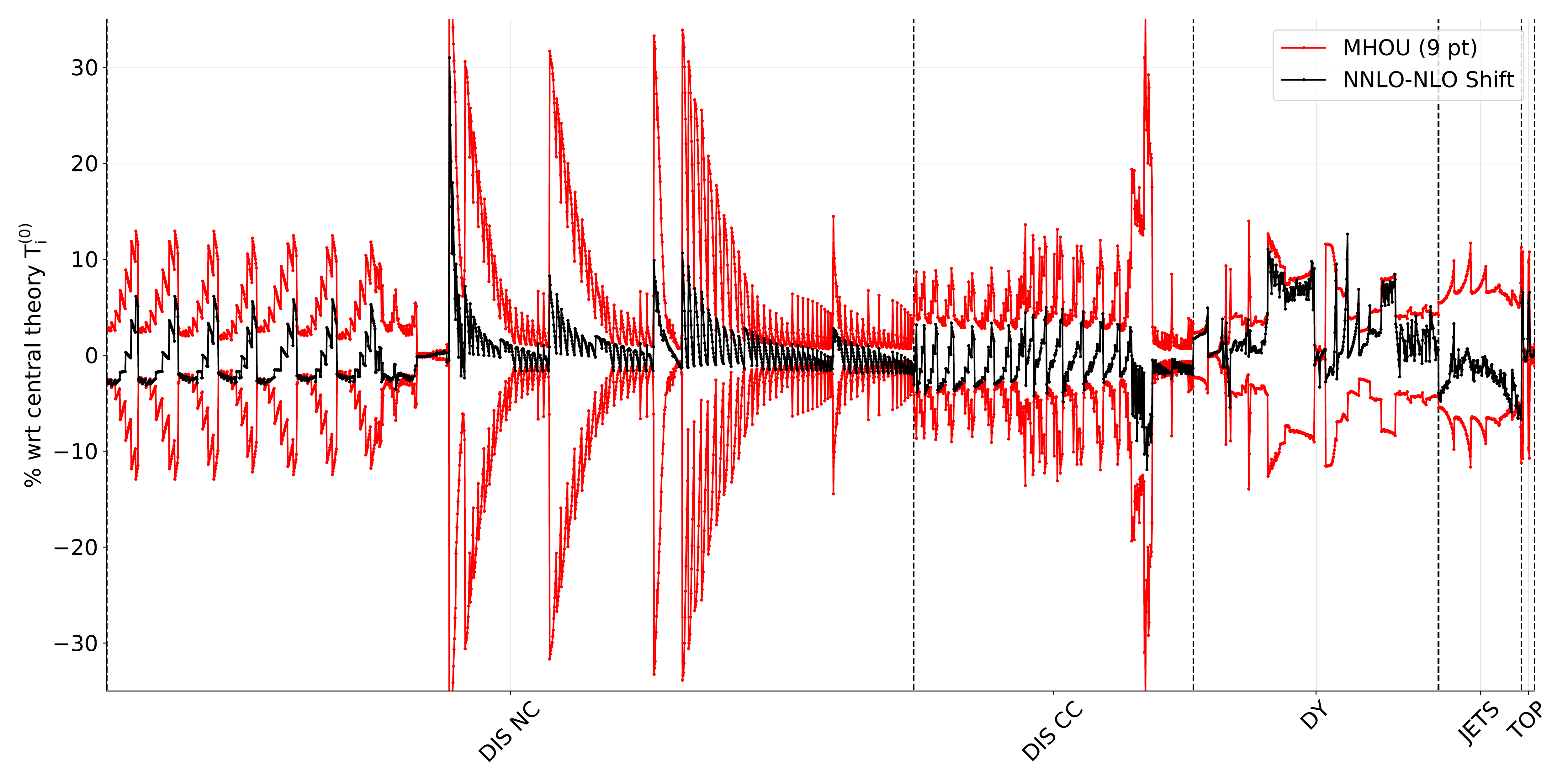}}
    \end{center}
  \vspace{-0.55cm}
  \caption{The relative uncertainties $\sigma_i$
    (9-point prescription) on
    the 2819 datapoints used in the PDF fit,
     compared to the known NLO-NNLO relative shifts $\delta_i$ in
     theory prediction.
  }
  \label{fig:shift_diag_cov_comparison}
\end{figure}

Since for the processes in Table~\ref{eq:expclassification} the
NNLO predictions are
known, we can validate the NLO covariance matrix against the known
NNLO results.
For this exercise, a common input NLO PDF is used in both cases.
In order to validate the diagonal elements of $S_{ij}$, which correspond to the
overall size of the MHOU, we first normalize it to the central theory prediction,
$\widehat{S}_{ij}=S_{ij}/T^{(0)}_iT^{(0)}_j$.
Then we compare
in Fig.~\ref{fig:shift_diag_cov_comparison} the relative uncertainties,
$\sigma_i=\sqrt{\widehat{S}_{ii}}$ to the relative shifts between predictions at
NLO and NNLO,
$\delta_{i}=(T^{(0),{\rm nnlo}}_{i}-T^{(0),{\rm nlo}}_{i})/T^{(0),{\rm nlo}}_i$,
for each of the $N_{\rm dat}=2819$ observables.
In all cases, $\delta_{i}$  turns out to be smaller or comparable
to $\sigma_i$, showing that this prescription provides a good
(if somewhat conservative) estimate of the diagonal theory uncertainties.

The validation of the full covariance matrix including correlations is
more subtle.
We first diagonalize $\widehat{S}_{ij}$, by finding the (orthonormal) eigenvectors $e^a_i$ which correspond to positive eigenvalues $(s^a)^2$: these
define a subspace $S$ orthonormal to the large null subspace.
The dimension $N_S$ of $S$
depends on the total number of independent scale variations,
the number of processes, and the correlation pattern. Its
determination is nontrivial, and it requires computing firstly the total
number of distinct scale variations for any pair of processes,
i.e., the total number of vectors $\Delta^{(k)}$ in
Eq.~(\ref{eq:covth}), and secondly determining the full
set of linear relations between them in
order to establish how many of them are independent (see Ref.~\cite{AbdulKhalek:2019ihb}).

For the 5 processes in
Table~\ref{eq:expclassification}, and the 9-point prescription, we find
$N_S=28$, while for the simpler 3-point prescription $N_S=6$.
We then compute the $N_S$ projections $\delta^a$
of the NLO-NNLO shifts $\delta_i$ along each eigenvector, and compare
them to the square root of the corresponding eigenvalues, $s^a$.
Finally we compute the length $|\delta^{\rm miss}_i|$ of the remaining
component of the vector $\delta_i$ that lies in the null subspace of
$\widehat{S}$.

  \begin{figure}[t]
    \begin{center}
    \makebox{\includegraphics[width=0.90\columnwidth]{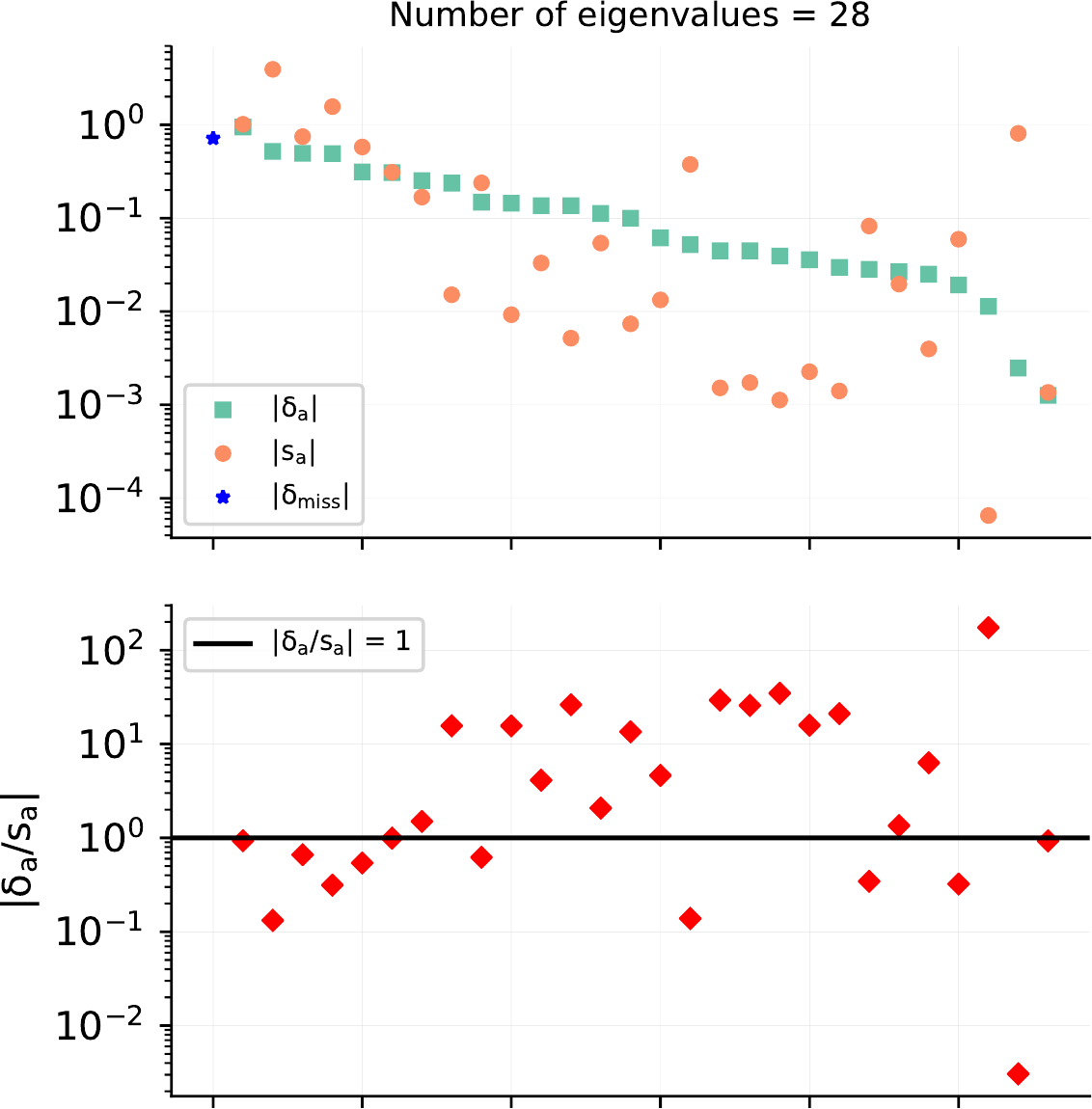}}
    \end{center}
  \vspace{-0.3cm}
  \caption{The square root eigenvalues $s^a$
    of the theory covariance
    matrix $\widehat{S}_{ij}$
    computed using the 9-point prescription, and the projections
    $\delta^a$ of the NNLO-NLO shift vector
    $\delta_i$ on the eigenvectors. The length
    $|\delta^{\rm miss}_i|$ of the component of $\delta_i$ lying in
    the null subspace of $\widehat{S}_{ij}$ is also shown.
  }
  \label{fig:eig_ratio}
\end{figure}

  The validation can be considered successful if the angle
  $\theta=\arcsin( |\delta^{\rm miss}_i|/|\delta_i|)$ is small,
meaning that the NNLO-NLO shift lies substantially within the
subspace $S$ estimated by the scale variations, and furthermore if $|\delta^a|\simeq|s^a|$, so that the size of the shift along each
eigenvector is correctly
estimated by the corresponding eigenvalue. Using the 9-point
prescription, for individual processes we
find $\theta=3^{\rm o}, 14^{\rm o}, 22^{\rm o}, 32^{\rm o}, 16^{\rm o}$ for top, jets, DY, NC and CC DIS respectively. For the complete dataset with the same prescription we
find $\theta=26^o$.

The projected shifts and
eigenvalues are compared in Fig.~\ref{fig:eig_ratio}. The size of the
eigenvalues generally falls as the projected shifts get smaller. For the six
largest eigenvectors the eigenvalue is always larger than the shift
and, in all but two cases, of very similar size to the shift. The 
seventh eigenvalue is smaller than, but of the same order as, the shift, 
while the eighth eigenvalue significantly underestimates the shift.  
However, given that
the eighth eigenvalue is already one order of magnitude smaller that
the first, this means that most of the shift is well described by the
theory covariance matrix, and somewhat overestimated by it in just a
few cases.
We conclude that the validation is successful: remarkably, the pattern of
correlations of theory shifts in a 2819-dimensional vector space is well
captured by just 28 nuisance parameters.

\begin{figure}[t]
    \begin{center}
    \makebox{\includegraphics[width=0.95\columnwidth]{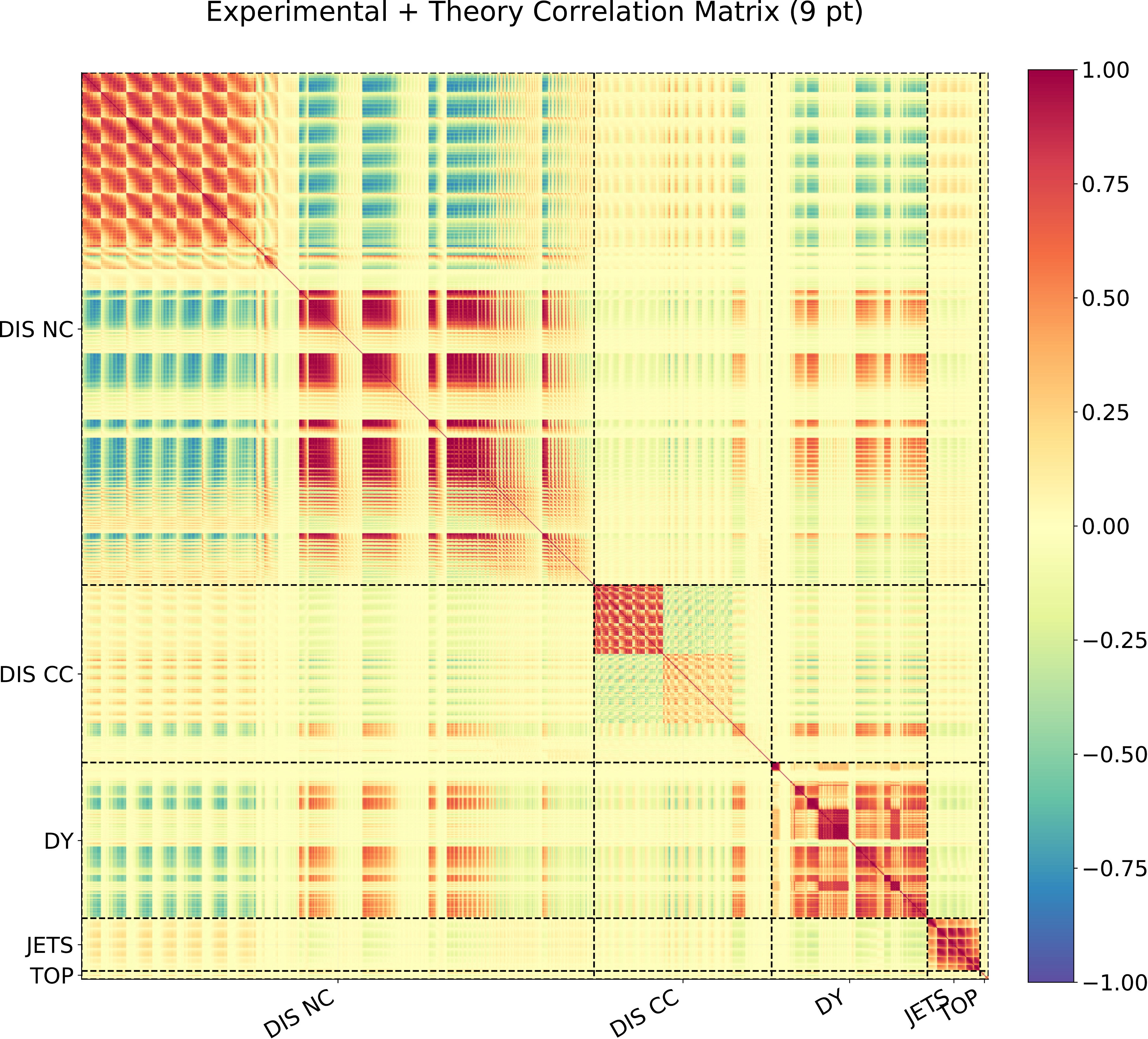}}
   \end{center}
  \vspace{-0.55cm}
  \caption{The combined experimental
    and theoretical (9-point) correlation matrix for the $N_{\rm dat}$ cross-sections
    in the fit.
  }
  \label{fig:default_theory0_plot_expcorrmat_heatmap}
\end{figure}

Adding the theory covariance matrix $S_{ij}$ to the experimental
covariance matrix $C_{ij}$, while increasing the diagonal uncertainty
on each individual prediction, also (and perhaps more importantly)
introduces a set of theory-induced
correlations between different experiments and processes, even when
the experimental data points are uncorrelated.
This is illustrated in
Fig.~\ref{fig:default_theory0_plot_expcorrmat_heatmap}, showing
the combined experimental and theoretical (9-point) correlation matrix: it
is clear that sizable correlations appear even between experimentally
unrelated measurements.

\begin{table}[t]
  \centering
  \renewcommand*{\arraystretch}{1.3}
  \begin{tabular}{|C{1.0cm}|C{1.6cm}|C{2.2cm}|C{2.2cm}|}
    \hline
    & $C$    &  $C+ S^{(\rm 3pt)}$ &  $C+ S^{(\rm 9pt)}$\\
    \hline
    $\chi^2$  & 1.139 &  1.139 & 1.109  \\
    $\phi$    & 0.314 &  0.394 & 0.415  \\
    \hline
  \end{tabular}
  \caption{\label{eq:chi2table} The
    central $\chi^2$ per datapoint and the average uncertainty
    reduction $\phi$ for the 3-point and 9-point fits.
  }
\end{table}

We can now proceed to a NLO global PDF determination with a theory covariance
matrix $S_{ij}$ computed using the 9-point prescription.
From the point of view of the NNPDF fitting methodology, the addition of
the theory contribution to the covariance matrix does not
entail any changes:
we follow the  procedure of Ref.~\cite{Ball:2014uwa},
but with the covariance matrix $C_{ij}$ now replaced by $C_{ij}+S_{ij}$,
both in the Monte Carlo replica generation and in the fitting.
In Table~\ref{eq:chi2table} we show some fit quality estimators
for the resulting PDF sets obtained using only the experimental
covariance matrix, alongside the theory covariance matrix
with two different prescriptions.

In particular, we show the
$\chi^2$ per datapoint
and the $\phi$ estimator~\cite{Ball:2014uwa}, which
gives the ratio of the uncertainty in the predictions using
the output PDFs to that of the original data, averaged in quadrature 
over all data.
The quality of the fit is improved by the inclusion
of the MHOU, with the 9-point prescription performing
rather better than 3-point.
Interestingly, $\phi$ only increases by around 30\% 
when one includes the theory covariance
matrix, much less than the 70\% 
one would expect taking into account the
relative size of the NLO MHOU and experimental uncertainties. This 
means that in the region of
the data, taking the MHOU into account increases the PDF
uncertainties only rather moderately. This suggests that the addition of
the MHOU is resolving some of the
tension between data and theory, so that the larger overall uncertainty is
partly compensated by the improved fit quality, though of course the
highly correlated nature of theory uncertainties also plays a role in
reducing their impact.

\begin{figure}[t]
  \begin{center}
    \makebox{\includegraphics[width=0.99\columnwidth]{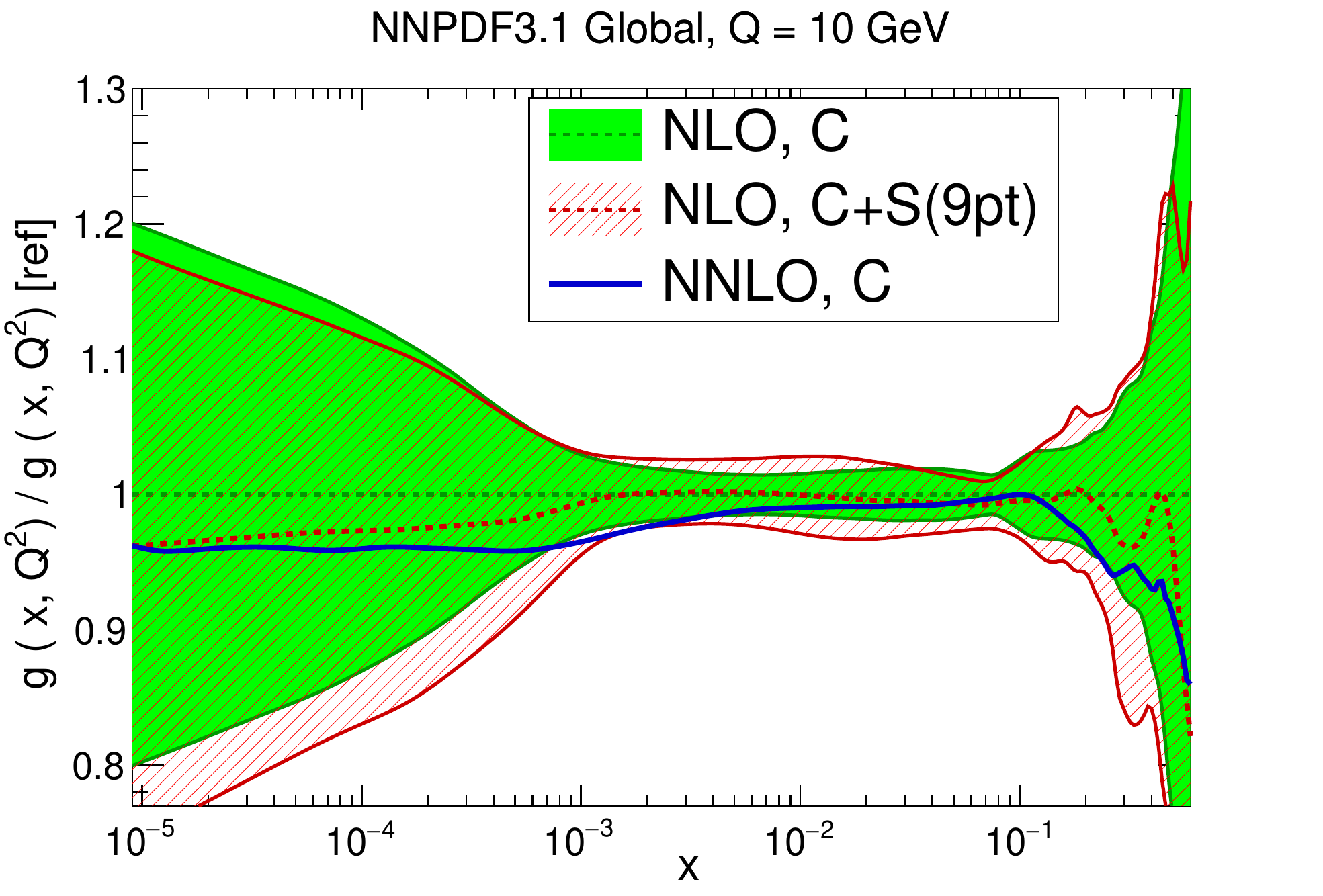}}
    \makebox{\includegraphics[width=0.99\columnwidth]{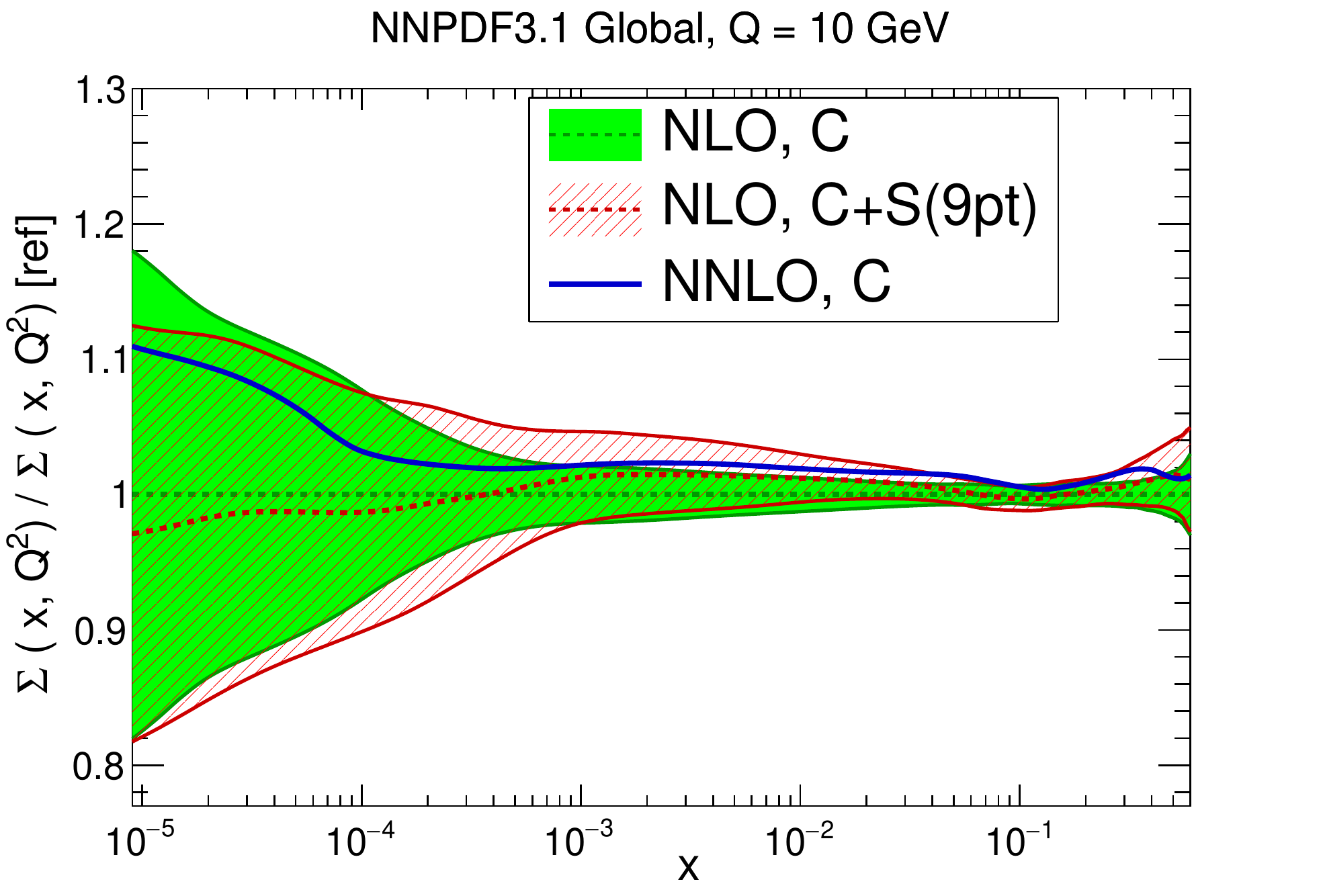}}
   \end{center}
  \vspace{-0.55cm}
  \caption{The gluon and quark singlet PDFs from the NNPDF3.1 NLO fits
    without and with the MHOU (9-points) in the
    covariance matrix at $Q=10$ GeV, normalized
    to the former.
    The central NNLO result is also shown.
  }
  \label{fig:Global-NLO-CovMatTH-EXP-vsTH}
\end{figure}

In Fig.~\ref{fig:Global-NLO-CovMatTH-EXP-vsTH}
we compare at $Q=10$ GeV the gluon and quark singlet PDFs obtained  at NLO with and without a
theory covariance matrix, normalized
to the latter. We also show the
central NNLO result when the theory covariance matrix is not included.
Three features of this comparison are apparent. First, when including
the MHOU, the increase in PDF uncertainty in the data region is quite
moderate, in agreement with the $\phi$ values of Tab.~\ref{eq:chi2table}.
Second, the NLO-NNLO shift is fully
compatible with the overall uncertainty.
Finally, the central value is also
modified by the inclusion of $S_{ij}$ in the fit, as the balance between different data sets adjusts according to their relative theoretical precision.
Interestingly, the central prediction shifts towards the known NNLO
result, showing that,
thanks to the inclusion of the MHOU, the overall fit quality has improved.

\begin{figure}[t]
  \begin{center}
    \makebox{\includegraphics[width=0.99\columnwidth]{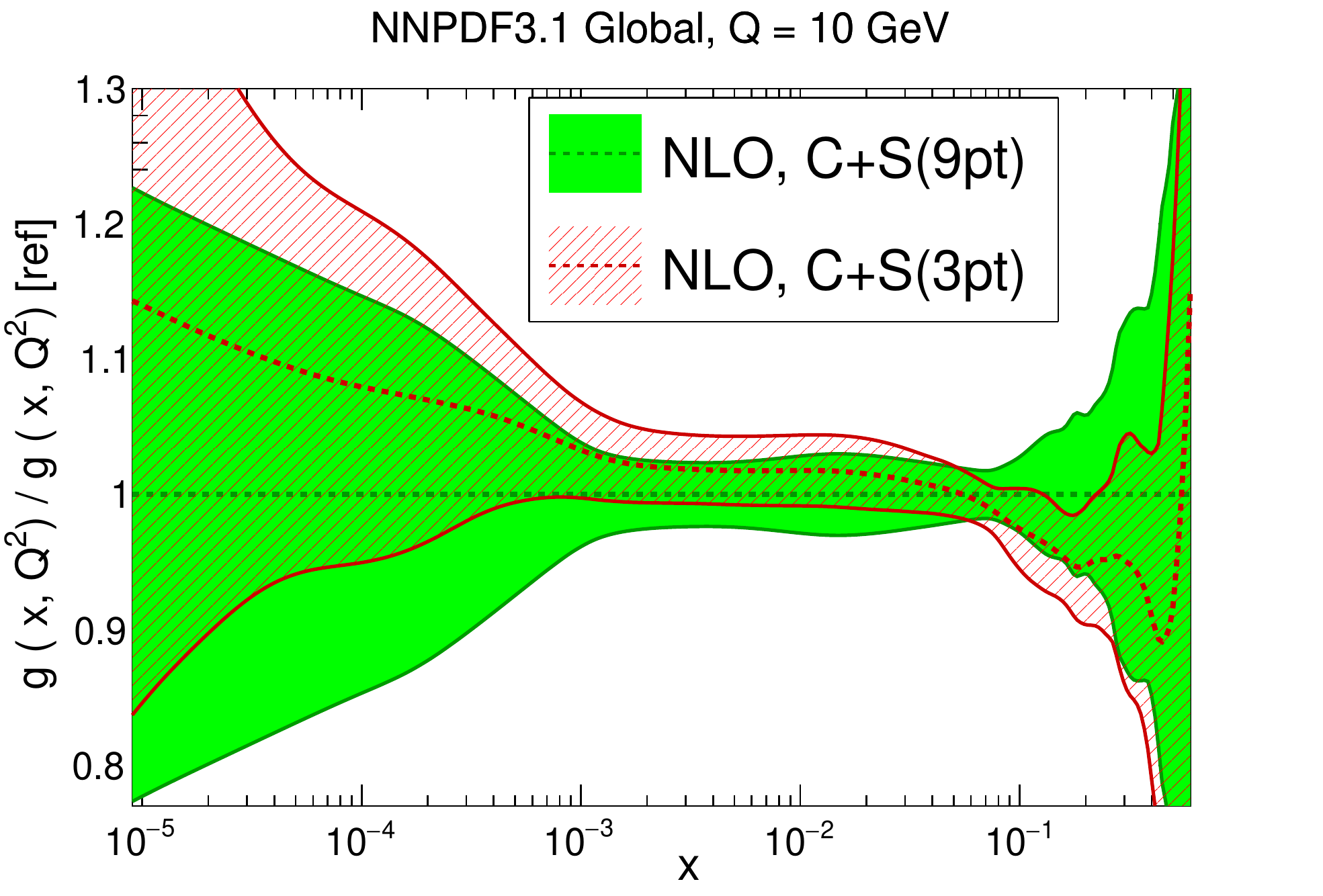}}
    \end{center}
  \vspace{-0.3cm}
  \caption{Same as Fig.~\ref{fig:Global-NLO-CovMatTH-EXP-vsTH} for the gluon,
    comparing the 3-point
    and 9-point prescriptions as a ratio to the latter.
   }
  \label{fig:Global-NLO-CovMatTH-prescriptions}
\end{figure}

Finally, in Fig.~\ref{fig:Global-NLO-CovMatTH-prescriptions}
we compare the dependence of the fit results on the specific choice of
prescription for $S_{ij}$, specifically for the  3-
and 9-point cases, normalized to the latter.
In general the two results are consistent, but
results with the 3-point prescription have
somewhat smaller uncertainties and, more importantly, their
central value is closer to
that when the MHOU is not included (see
Fig.~\ref{fig:Global-NLO-CovMatTH-EXP-vsTH}), so that the improved
agreement between the NLO and full NNLO noted in
Fig.~\ref{fig:Global-NLO-CovMatTH-EXP-vsTH} would be mostly lost if
the 3-point prescription were adopted, providing further
confirmation for preferring the 9-point prescription.

It is important to understand that the meaning of PDFs and their
uncertainties
changes once the theory covariance matrix is included: so the error
bands e.g. in Fig.~\ref{fig:Global-NLO-CovMatTH-EXP-vsTH} have a
different meaning according to whether the theory covariance matrix is
included. When it is included, PDF uncertainties account for data
and methodological uncertainties, but also for MHOUs. Also, 
their central values now optimize the agreement with data based on a
$\chi^2$ which includes MHOUs.

The usage of these PDFs is accordingly
different. Firstly, they should be combined with hard
cross-sections which also include MHOU. The MHOU on the prediction
and the PDF uncertainty (now also including MHOUs) should be combined
in the standard way (i.e. in quadrature), since with a universal PDF it 
is not possible to keep
track of the correlations (which surely exist) between MHOU in
processes used for PDF determination, and the MHOU in the prediction
itself. This neglected correlation is likely to be a small effect in
most situations~\cite{AbdulKhalek:2019ihb}, and it leads to a conservative
uncertainty estimate. Second, it is important to keep in mind that
MHOUs in the theory prediction
must be included in the computation of the $\chi^2$ when
assessing the agreement of these PDFs with new data, since, as we have
seen, their central value is shifted as a consequence of the inclusion
of the MHOUs.

In summary, we have presented the first global PDF analysis
that accounts for the MHOU
associated with the fixed order QCD perturbative calculations used in the fit.
The inclusion of the MHOU shifts central values by an amount that is
not negligible on the scale of the PDF uncertainty,
moving the NLO result towards the NNLO result.
PDF uncertainties increase moderately,
because of the improvement of fit quality due to the rebalancing of datasets
according to their theoretical precision. 
For this to be effective, the correlations in $S_{ij}$ play a crucial
role. These correlations are rather more extensive than those related
to experimental systematics, since all different measurements of the
same process are correlated through their common MHO
corrections, and different processes are correlated
through MHO corrections to perturbative evolution. A
more accurate treatment of these correlations (especially those
related to perturbative evolution) will be the subject of future studies.

Our results  pave the way towards a fully consistent treatment of MHOU
for precision LHC phenomenology. The NLO results presented here
will be upgraded to NNLO, facilitated
by tools such as the  {\tt APPLfast} grid interface
to the {\tt NNLOJET} program~\cite{Gehrmann:2018szu}. We thus anticipate that
the upcoming  NNPDF4.0 PDF set will be able to fully account for MHOU both
at NLO and NNLO, as well as other sources of theory uncertainty,
such as those related to nuclear corrections~\cite{Ball:2018twp,AbdulKhalek:2019mzd}.

\paragraph{Acknowledgments.}
R.~D.~B. is supported by the UK Science and Technology Facility Council through grant ST/P000630/1.\sloppy
S.~F. is supported by the European Research Council under
the European Union's Horizon 2020 research and innovation Programme
(grant agreement n.740006).
T.~G. is supported by The Scottish Funding Council,
grant H14027.
Z.~K. is supported by the European Research Council Consolidator Grant
``NNLOforLHC2''.
E.~R.~N. is supported by the European Commission through the
Marie Sk\l odowska-Curie Action ParDHonS\_FFs.TMDs (grant number 752748).
R.~L.~P. and M.~W. are supported by the STFC grant ST/R504737/1.
J.~R. is supported by the European Research Council Starting
Grant ``PDF4BSM'' and by the Netherlands Organization for Scientific
Research (NWO).
L.~R. is supported by the European Research Council Starting
Grant ``REINVENT'' (grant number 714788).
M.~U. is partially supported by the STFC grant ST/L000385/1 and funded by the Royal Society grants DH150088 and RGF/EA/180148.
C.~V. is supported by the STFC grant ST/R504671/1.

\bibliographystyle{spphys}
\bibliography{bib/nnpdf31_therr_covmat}

\begin{thebibliography}{10}
\providecommand{\url}[1]{{#1}}
\providecommand{\urlprefix}{URL }
\expandafter\ifx\csname urlstyle\endcsname\relax
  \providecommand{\doi}[1]{DOI \discretionary{}{}{}#1}\else
  \providecommand{\doi}{DOI \discretionary{}{}{}\begingroup
  \urlstyle{rm}\Url}\fi

\bibitem{deFlorian:2016spz}
D.~de~Florian, et~al.,   (2016), 1610.07922

\bibitem{Cepeda:2019klc}
M.~Cepeda, et~al.,   (2019), 1902.00134

\bibitem{Gao:2017yyd}
J.~Gao, L.~Harland-Lang, J.~Rojo, Phys. Rept. \textbf{742}, 1 (2018),
  1709.04922

\bibitem{Rojo:2015acz}
J.~Rojo, et~al., J. Phys. \textbf{G42}, 103103 (2015), 1507.00556

\bibitem{Ball:2017nwa}
R.D. Ball, et~al., Eur. Phys. J. \textbf{C77}(10), 663 (2017), 1706.00428

\bibitem{Cacciari:2011ze}
M.~Cacciari, N.~Houdeau, JHEP \textbf{1109}, 039 (2011), 1105.5152

\bibitem{David:2013gaa}
A.~David, G.~Passarino, Phys. Lett. \textbf{B726}, 266 (2013), 1307.1843

\bibitem{Bagnaschi:2014wea}
E.~Bagnaschi, M.~Cacciari, A.~Guffanti, L.~Jenniches, JHEP \textbf{02}, 133
  (2015), 1409.5036

\bibitem{Ball:2018odr}
R.D. Ball, A.~Deshpande,  (2018). 1801.04842

\bibitem{Ball:2018twp}
R.D. Ball, E.R. Nocera, R.L. Pearson, Eur. Phys. J. \textbf{C79}(3), 282
  (2019), 1812.09074

\bibitem{Pearson:2018tim}
R.L. Pearson, C.~Voisey, Nucl. Part. Phys. Proc. \textbf{300-302}, 24 (2018),
  1810.01996

\bibitem{AbdulKhalek:2019ihb}
R.~Abdul~Khalek, et~al.,   (2019), 1906.10698

\bibitem{Ball:2018iqk}
R.D. Ball, S.~Carrazza, L.~Del~Debbio, S.~Forte, Z.~Kassabov, J.~Rojo,
  E.~Slade, M.~Ubiali, Eur. Phys. J. \textbf{C78}(5), 408 (2018), 1802.03398

\bibitem{zahari_kassabov_2019_2571601}
Z.~Kassabov.
\newblock {Reportengine: A framework for declarative data analysis}.
\newblock https://doi.org/10.5281/zenodo.2571601 (2019)

\bibitem{Bertone:2013vaa}
V.~Bertone, S.~Carrazza, J.~Rojo, Comput.Phys.Commun. \textbf{185}, 1647
  (2014), 1310.1394

\bibitem{Bertone:2016lga}
V.~Bertone, S.~Carrazza, N.P. Hartland, Comput. Phys. Commun. \textbf{212}, 205
  (2017), 1605.02070

\bibitem{Carli:2010rw}
T.~Carli, et~al., Eur.Phys.J. \textbf{C66}, 503 (2010), 0911.2985

\bibitem{Ball:2014uwa}
R.D. Ball, et~al., JHEP \textbf{04}, 040 (2015), 1410.8849

\bibitem{Gehrmann:2018szu}
T.~Gehrmann, et~al., PoS \textbf{RADCOR2017}, 074 (2018), 1801.06415

\bibitem{AbdulKhalek:2019mzd}
R.~Abdul~Khalek, J.J. Ethier, J.~Rojo,   (2019), 1904.00018

\end{thebibliography}

\end{document}